\documentclass[prd,aps,twocolumn,nofootinbib]{revtex4}
\usepackage{epsfig}
\usepackage{amsfonts}
\usepackage{rotating}
\usepackage{sparticles}
\usepackage{dcolumn}
\usepackage{bm}
\usepackage{hyperref}
\hypersetup{
    bookmarks=true,         
    unicode=false,          
    pdftoolbar=true,        
    pdfmenubar=true,        
    pdffitwindow=false,     
    pdfstartview={FitH},    
    pdftitle={My title},    
    pdfauthor={Author},     
    pdfsubject={Subject},   
    pdfcreator={Creator},   
    pdfproducer={Producer}, 
    pdfkeywords={keyword1} {key2} {key3}, 
    pdfnewwindow=true,      
    colorlinks=true,       
    linkcolor=red,          
    citecolor=cyan,        
    filecolor=magenta,      
    urlcolor=green,           
    linktocpage=true
}
\usepackage{amsmath,amssymb,amsthm,graphicx,latexsym}
\usepackage{subfigure}
\usepackage{pstricks}
\usepackage{color}

\usepackage{mathrsfs}

\newcommand{\be}{\begin{equation}}
\newcommand{\ee}{\end{equation}}
\newcommand{\bea}{\begin{eqnarray}}
\newcommand{\eea}{\end{eqnarray}}
\newcommand{\bml}{\begin{subequations}}
\newcommand{\eml}{\end{subequations}}
\newcommand{\bfig}{\begin{figure}}
\newcommand{\efig}{\end{figure}}

\newcommand{\slashed}{\hspace{-1.1ex}/}

\begin{document}

\title{Primordial blackholes  and gravitational waves for an inflection-point model of inflation}
\author{Sayantan Choudhury $^{1}$  and Anupam Mazumdar $^{2}$}
\affiliation{$^1$Physics and Applied Mathematics Unit, Indian Statistical Institute, 203 B.T. Road, Kolkata 700 108, INDIA}
\affiliation{$^2$Consortium for Fundamental Physics, Physics Department, Lancaster University, LA1 4YB, UK}

\begin{abstract}
In this article we provide a new {\it closed} relationship between cosmic abundance of primordial gravitational waves and 
primordial blackholes originated from initial inflationary perturbations for {\it inflection-point} models of inflation where inflation 
occurs below the Planck scale. The current Planck constraint on tensor-to-scalar ratio, running of the spectral tilt, and from the abundance of 
dark matter content in the universe, we can deduce a strict bound on the current abundance of primordial blackholes to be
within a range, $9.99712\times 10^{-3}<\Omega_{PBH}h^{2}<9.99736\times 10^{-3}$.
\end{abstract}


\maketitle
\section{Introduction} 

In the Einstein's general relativity (GR) the primordial blackholes (PBHs) with a small mass can be created during the radiation 
epoch due to over density on length scales $R\sim 1/k_{PBH}$, which is typically much smaller than the 
pivot scale at which the relevant perturbations re-enter the Hubble patch for the large scale structures, $k_* $~\cite{Hawking:1971ei,Carr,Carr:1994ar}.
Typically the regions with a mass less than the size of the Hubble radius can collapse to form PBHs, i.e. 
$M\leq \gamma M_H\sim \gamma (4\pi/3)\rho H^{-3}(t)\approx 2\times 10^{5}\gamma(t/1~{\rm s})M_{\odot}$, where 
$\rho$ is the energy density of the radiation epoch, $H$ is he Hubble radius, $M_{\odot}\sim 10^{33}$g, and $\gamma\leq 0.2$
is the numerical factor during the radiation era which depends on the dynamics of gravitational collapse~\cite{Carr}.
For instance, an economical way would be to create PBH abundance  from an initial primordial inflationary 
fluctuations which had already entered the Hubble patch during the radiation era, but whose amplitude had increased on small scales due to the 
running in the spectral index, $n_s$~\cite{Lyth:2005ze,Manuel:2011}~\footnote{A word of caution - GR is not an ultraviolet (UV) complete theory. An UV completion of gravity may naturally lead to ghost free and asymptotically free theory of gravity, as recently proposed in Ref.~\cite{Biswas:2011ar,Biswas:2013ds}. In such a class of theory it has been shown that mini-blackhole with a mass less than the Planck mass, i.e. $10^{-5}$g does not have a singularity and nor does it have a horizon~\cite{,Biswas:2011ar}. }.

An interesting observation was first made in Ref.~\cite{Hotchkiss:2011gz} and
 in Refs.~\cite{Choudhury:2013jya,Choudhury:2013iaa,Choudhury:2014sxa,Choudhury:2014uxa,Choudhury:2014kma,Choudhury:2014wsa}, 
that a sub-Planckian inflaton field can create a significant primordial gravitational waves (PGWs) provided the last $50-60$ e-foldings of inflation is driven 
through the {\it inflection-point}, such that the tensor to scalar ratio saturates the 
Planck constrain, $r\leq 0.12$~\cite{Ade:2013uln}. One requires a marginal running in the power spectrum which is now well constrained by the Planck+WMAP9 combined data.
A valid particle physics model of inflation can only occur below the cut-off scale of gravity,
see for a review on particle physics models of inflation~\cite{Mazumdar:2010sa},  it would be interesting to study the implications of the running of 
the spectral tilt, $\alpha_s$, for both PGWs and PBHs. 

Formation of the significant amount of PBHs on a specific mass scale is realized iff
 the power spectrum of primordial fluctuations has amplitude
$10^{-2}-10^{-1}$ on the corresponding scales \cite{Yokoyama:1998pt}. In such a
physical situation the second-order effects in the cosmological perturbation are expected to play
an significant role in the present set up. Also such non-negligible effects also generate tensor fluctuations
to produce PGWs from scalar-tensor modes via terms
quadratic in the first-order matter and metric perturbations \cite{Saito:2009,Ananda:2006af,Matarrese:1993zf}. Most importantly, their
amplitude may well exceed the first-order tensor perturbation generated by quantum fluctuation during inflation 
in the present set up as the amplitude of density fluctuations required to produce PBHs is large.

The aim of this paper is to provide an unique link between the current abundance of PBHs, $\Omega_{PBH}(\eta_0)=\rho_{PBH}/\rho_c$, and
the abundance of primordial gravitational waves $\Omega_{GW}=\rho_{GW}/\rho_c$ in our universe originated from the primordial fluctuations, where
$\eta_0$ is the present conformal time and $\rho_c$ denotes the critical energy density of the universe. With the help of Planck data, we will be able to constrain 
a concrete bound on $\Omega_{PBH}h^2$.

\section{PBH formation} 


\begin{figure*}[t]
\centering
\subfigure[]{
    \includegraphics[width=8.1cm, height=6.5cm] {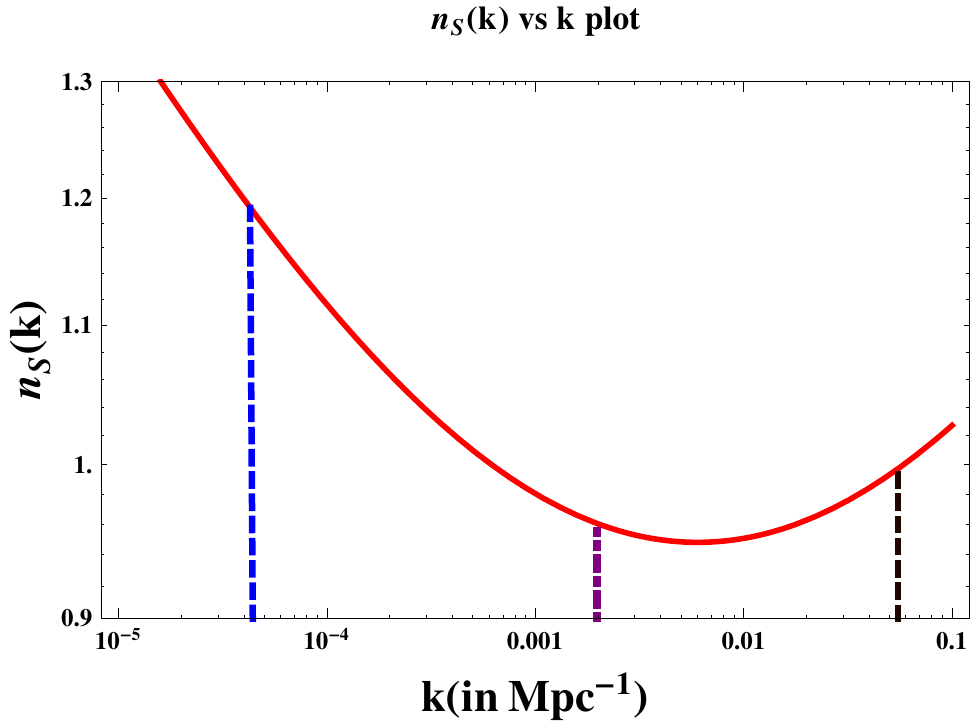}
    \label{fig:subfig1}
}
\subfigure[]{
    \includegraphics[width=8.1cm, height=6.5cm] {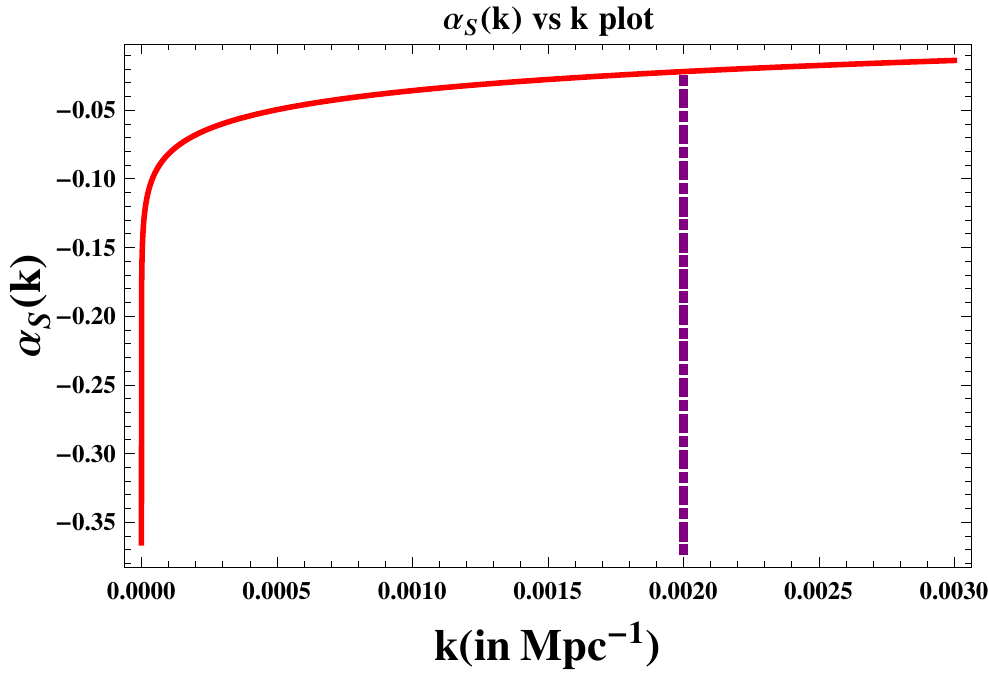}
    \label{fig:subfig2}
}
\caption[Optional caption for list of figures]{ In \ref{fig:subfig1}, we show the scalar spectral index $n_{s}(k)$,
 and in \ref{fig:subfig2}, we show the running of the scalar spectral index $\alpha_{s}(k)$, with respect to the momentum scale $k$.
The {\bf black} dotted line corresponds to $k_{\Lambda}=0.056~{\rm Mpc}^{-1}$ for $l_{\Lambda}=2500$, the \textcolor{blue}{\bf blue} dotted line corresponds to
$k_{L}=4.488\times 10^{-5}~{\rm Mpc}^{-1}$ for $l_{L}=2$,  and in all the plots \textcolor{violet}{\bf violet} dashed dotted line
 represents the pivot scale of momentum at $k_{\star}=0.002~{\rm Mpc}^{-1}$ for $l_{\star}\sim 80$ at which
 $P_{S}(k_{\star})=2.2\times 10^{-9}$, $n_{S}(k_{\star})=0.96$, $\alpha_{S}(k_{\star})=-0.02$. 
In both fig~\ref{fig:subfig1} and fig~\ref{fig:subfig2} the PBH formation scale $k_{PBH}$ is lying within the region bounded by the \textcolor{blue}{\bf blue} dotted line 
and the \textcolor{violet}{\bf violet} dashed dotted line.
}
\label{fig1}
\end{figure*}

Let us first start the discussion with the amplitude of the scalar power spectrum, which
 is defined at any arbitrary momentum scale lying within the window, $k_{L}<k<k_{\Lambda}$, by:
\be\begin{array}{llll}\label{dx2} 
\displaystyle 
P_{s}(k)=P_{s}(k_{*})\left(\frac{k}{k_{*}}\right)^{n_{s}-1
+\frac{\alpha_{s}}{2}\ln\left(\frac{k}{k_{*}}\right)+\frac{\kappa_{s}}{6}\ln^{2}\left(\frac{k}{k_{*}}\right)+\cdots},\end{array}\ee
where the parameters $n_{s}$, $\alpha_{s}$ and 
$\kappa_{s}$ are spectral tilt, running and running of running of the tilt
of the scalar perturbations defined in the momentum pivot scale $k_{*}$.
 Also note that the PBH formation scale is lying within the window, $k_{L}<k_{PBH}<k_{*}<k_{\Lambda}$. In the realistic situations
the upper and lower bound of the momentum scale is fixed at, $k_{\Lambda}=0.056~{\rm Mpc}^{-1}$
 for $l_{\Lambda}=2500$ and $k_{L}=4.488\times 10^{-5}~{\rm Mpc}^{-1}$ for $l_{L}=2$.

Within this window,  we need to modify the power law parameterization of the power spectrum by incorporating the effects of higher order Logarithmic corrections in terms
of the non-negligible running, and running of the running of the spectral tilt as shown in Eq(\ref{dx2}), which involves higher order slow-roll corrections 
in the next to leading order of effective field theory of inflation. This is important to consider since PlanckWMAP9 combined data have already placed interesting constraints on 
$n_{s}(k_*)=0.9603\pm 0.0073$ (within $2\sigma$ C.L.), $\alpha_{s}(k_*)=−0.0134 \pm 0.0090$, and $\kappa_{s}(k_*)= 0.020^{+0.016}_{-0.015}$ 
(within $1\sigma-1.5\sigma$ C.L.)~\cite{Ade:2013uln}.

Further using Eq~(\ref{dx2}), spectral tilt, running of the tilt, and running of the running of the tilt for the scalar perturbations can be written 
at any arbitrary momentum scale within the widow, $k_{l=2}<k<k_{l=2500}$,
as:
\be\begin{array}{llll}\label{tilt}
    \displaystyle n_{s}(k)=1+\frac{d\ln P_{s}(k)}{d\ln k}\\ \displaystyle~~~~~~~~=n_{s}(k_{*})+\alpha_{s}(k_{*})\ln\left(\frac{k}{k_{*}}\right)
\\~~~~~~~~~~~~~~~~~~~~~~~~~~\displaystyle +\frac{\kappa_{s}(k_{*})}{2}\ln^{2}\left(\frac{k}{k_{*}}\right)+\cdots\,
   \end{array}\ee
\be\begin{array}{llll}\label{run1}
    \displaystyle \alpha_{s}(k)=\frac{dn_{s}(k)}{d\ln k}=\alpha_{s}(k_{*})
+\kappa_{S}(k_{*})\ln\left(\frac{k}{k_{*}}\right)+\cdots\,
   \end{array}\ee
\be\begin{array}{llll}\label{run2}
    \displaystyle \kappa_{s}(k)=\frac{d\alpha_{s}(k)}{d\ln k}=\frac{d^2n_{s}(k)}{d\ln k^{2}}\approx\kappa_{s}(k_{*})+\cdots\,.
   \end{array}\ee
At the scale of PBH formation, $k=k_{PBH}$, the value of the tilt, and running of the running of tilt for the scalar perturbations can be
expanded around the  pivot scale ($k_{*}$) as :
\be\begin{array}{llll}\label{e1}
    \displaystyle n_{s}(k_{PBH})=n_{s}(k_{*})+\alpha_{s}(k_{*})\ln\left(\frac{k_{PBH}}{k_{*}}\right)
\\~~~~~~~~~~~~~~~~~~~~~~~~~~\displaystyle +\frac{\kappa_{s}(k_{*})}{2}\ln^{2}\left(\frac{k_{PBH}}{k_{*}}\right)+\cdots\,
   \end{array}\ee
\be\begin{array}{llll}\label{e11}
    \displaystyle \alpha_{s}(k_{PBH})=\alpha_{s}(k_{*})
+\kappa_{S}(k_{*})\ln\left(\frac{k_{PBH}}{k_{*}}\right)+\cdots\,
   \end{array}\ee
\be\begin{array}{llll}\label{e111}
    \displaystyle \kappa_{s}(k_{PBH})\approx\kappa_{s}(k_{*})+\cdots\,.
   \end{array}\ee
 provided the expansion is valid when, $k_{PBH}<k_{*}$. Throughout the article we will fix the pivot scale to be the same as that of the Planck, 
 $k_{*}\sim 0.002~{\rm Mpc}^{-1}$. However, it is important to note that the physics is independent
 of the choice of the numerical value of $k_{*}$. The $\cdots$ represent higher order slow-roll corrections appearing in the expansion. 
Here the pivot scale of momentum $k_{*}$ is a normalization scale and is of the order 
of the UV regularized scale of the momentum cut-off of the power spectrum beyond which the 
 logarithmically corrected power law parameterization of the primordial power spectrum for scalar modes does not hold good.
More precisely, $k_{*}$ be a floating momenta in the present context.
Additionally, we have used another restriction on the momentum scale, $k_{l=2}<k_{PBH}<k_{*}<k_{l=2500}$. For more details see Eq~(\ref{dx2}) mentioned later. 
The initial PBHs mass, ${M}_{PBH}$, is related to the Hubble mass, ${M}$, by:
\be\begin{array}{llll}\label{d31}
\displaystyle 
{M}_{PBH}={M }\gamma= \frac{4\pi}{3}\gamma \rho  {\cal H}^{-3}\,,\end{array}\ee
at the Hubble entry, where the Hubble parameter ${\cal H}$ is defined in terms of  the conformal time, $\eta$.
The PBH is formed when the density
fluctuation exceeds the threshold for PBH formation given by the {\it Press--Schechter theory}~\cite{Press:1974}
\be\begin{array}{lll}\label{fofm}
\displaystyle  f( \geq{M}) = 2 \gamma \int _{\varTheta_{\rm th}}^{\infty} {\cal\text{d}\varTheta\,}{\cal P}(\varTheta; M(k_{PBH}))\,.
\end{array}\ee
Here ${\cal P}(\varTheta; M(k_{PBH}))$ is the {\it Gaussian probability distribution function} of the
linearized density field $\varTheta$ smoothed on a scale, $k_{PBH}=1/R$, by~\cite{Green:2004}:
\be\begin{array}{llll}\label{cfr1}
\displaystyle 
{\cal P}(\varTheta;k_{PBH}) = \frac {1} {\sqrt{2\pi} \varSigma_{\varTheta}({k_{PBH}})} \exp\left( -\frac
{\varTheta^{2}} {2\varSigma_{\varTheta}^{2}({k_{PBH}})} \right)\end{array}\ee
where the standard deviation is given by
\be\begin{array}{lll} \label{sigma}
\displaystyle\varSigma_{\varTheta}({ k_{PBH}}) =\sqrt{ \int_{0}^{\infty}\dfrac{\text{d}k} {k} \exp\left(-\frac{k^{2}}{k^{2}_{PBH}} \right)\ {P}_{\varTheta}(k)
}\,.
\end{array}\ee
Here it is important to note that the fine details of our conclusions might change while taking into N body simulations
into account \cite{Lacey:1994su,springel,Navarro:1996gj,Dayal:2012ah}. For a generic class of inflationary models, linearized smooth density 
 field $\varTheta(k)$, and the corresponding power spectrum ${P}_{\varTheta}(k)$
can be written as :
\be\begin{array}{ll}\label{psx1} 
\displaystyle 
\varTheta(k)=\frac{2}{5}\left( \frac{k}{a{\cal H}}
\right)^2{\cal R}_{c}(k),\\
\displaystyle 
{P}_{\varTheta}(k)
=\frac{4} {25}\frac{(1+w)^2} {\left(1+\frac{3}{5}w\right)^2}\left( \frac{k}{a{\cal H}}
\right)^4 {P}_{S}(k),
\end{array}\ee
where $w$ represents the effective equation of state parameter after the end of inflation. Assuming that the inflaton 
decays into the relativistic species {\it instantly}, we may be able to fix $w\approx 1/3$, for a radiation dominated universe.
Additionally, ${\cal R}_{c}(k)$ characterizes the curvature perturbation, and 
$P_S$ denotes the amplitude of the scalar power spectrum. 

Now substituting Eq.~(\ref{psx1}) and Eq.~(\ref{dx2}) in Eq.~(\ref{sigma}),
for $k_{PBH}=1/R$,  we can express $\varSigma_{\varTheta}(k_{PBH})$ as:
\begin{widetext}\be\begin{array}{llll}\label{sadt}
\displaystyle 
\varSigma_{\varTheta}(k_{PBH}) =\frac{2}{5}
\frac{(1+w)\sqrt{P_{S}(k_{*})}} {\left(1+\frac{3}{5}w\right)}\left(\frac{k_{*}}{a{\cal H}}\right)^{2}
\sqrt{ \int_{k_{L}}^{k_{\Lambda}}\frac{dk}{k_{*}}\exp\left(-\frac{k^{2}}{k^{2}_{PBH}} \right)\ \left(\frac{k}{k_{*}}\right)^{n_{s}+2
+\frac{\alpha_{s}}{2}\ln\left(\frac{k}{k_{*}}\right)+\frac{\kappa_{s}}{6}\ln^{2}\left(\frac{k}{k_{*}}\right)+\cdots}
}\,\\
\displaystyle
~~~~~~~~~~~~~~=\frac{1}{5}
\frac{(1+w)\sqrt{P_{S}(k_{*})}} {\left(1+\frac{3}{5}w\right)}\left(\frac{k_{*}}{a{\cal H}}\right)^{2}\sqrt{A+Bn_{s}(k_{*})+C\alpha_{s}(k_{*})
+D\kappa_{s}(k_{*})+\cdots}\end{array}\ee
\end{widetext}
where we have reparametrized the integral in terms of the {\it regulated} UV (high) and IR (low) momentum scales. The cut-offs ($k_{\Lambda}$ and $k_{L}$) 
are floating momenta to collect only the finite contributions. The technique we imploy here has a similarity to the cut-off regularization scheme, which 
is being introduced in such a fashion that after
taking the physical limit, ($k_{\Lambda}\rightarrow \infty$, $k_{L}\rightarrow 0$), the result returns to the original infinite integral.
 
Here the UV and  the IR cut-offs must satisfy the constraint condition, $k_{L}\ll k_{PBH}\ll k_{*}\ll k_{\Lambda}$, for which the integral 
appearing in the expression for the standard deviation can be
regularized. In Eq.~(\ref{sadt}), $A,~B,~C$ and $D$ are all momentum dependent coefficients which are explicitly mentioned in the appendix, see 
Eq.~(\ref{coeff}). Moreover, at the Hubble exit an additional constraint, $k_{*}=a{\cal H}$,
will have to be satisfied in order to do the matching of the long and short wavelength perturbations. 

Hence, substituting the explicit expressions for $P_{S}$, $n_{s}$, $\alpha_{s}$ and $\kappa_{s}$ in presence of the
higher order corrections at the pivot scale $k_{*}$, the simplified expression for the regularized standard deviation in terms of the leading order 
slow-roll parameters can be written as:
%
\be\begin{array}{llll}\label{sadtc1}
\displaystyle 
\varSigma_{\varTheta}(k_{PBH}) =
\frac{(1+w)\sqrt{\frac{A V_{*}}{\epsilon_{V}(k_{*})}}} {8\sqrt{6}\pi M^{2}_{pl}
\left(1+\frac{3}{5}w\right)}
\left\{1+\frac{B}{A}\left({\cal C}_{E}+\frac{2}{5}\right)\eta_{V}(k_{*})
\right.\\ \left.~~~~~~~~~~~~~~~~~~~~~~~~\displaystyle
-\frac{B}{A}\left(2{\cal C}_{E}
+\frac{11}{5}\right)\epsilon_{V}(k_{*})-\frac{C}{5A}
\xi^{2}_{V}(k_{*})
\right.\\ \left.~~~~~~~~~~~~~~~~~~~~~~~~~~~~~~~~~\displaystyle+\frac{2D}{5A}
\sigma_{V}^{3}(k_{*})
+\cdots\right\}\,\end{array}\ee
%
where ${\cal C}_{E}=4(\ln 2+\gamma_{E})-5$, and $\gamma_{E}=0.5772$ is the {\it Euler-Mascheroni constant}.
Here the $(\epsilon_{V},\eta_{V},\xi^{2}_{V},\sigma^{3}_{V})$ are slow roll parameters for a given inflationary potential $V(\phi)$. 
It is important to mention that the results obtained in this paper are inflation centric - true only for {\it inflection point models} of inflation.

\section{PBH and GW for sub-Planckian model of inflation} 

For a successful inflation, the potential should be flat enough, and for a generic 
inflationary potential around the vicinity of the VEV $\phi_0$, where inflation occurs, we may impose 
the flatness condition such that, $V^{\prime\prime}(\phi_0)\approx 0$. This yields a simple flat potential which has been 
imposed in many well motivated particle physics models of inflation with an {\it inflection-point}~\cite{Enqvist:2010vd}:
\begin{equation}
V(\phi)=\alpha+\beta(\phi-\phi_{0})+\gamma(\phi-\phi_{0})^{3}+\kappa(\phi-\phi_{0})^{4}+\cdots \,,
\end{equation}
where $\alpha$ denotes the height of the potential, and the coefficients $\beta,~\gamma,~\kappa$ determine the shape of the 
potential in terms of the model parameters. Note that at this point, we do not need to specify any particular model of inflation
for the above expansion of $V(\phi)$. However, not all of the coefficients are independent once we prescribe the
model of inflation here. This only happens if the VEV of the inflaton $\phi_0<M_{p}$ must be bounded by the cut-off of the particle
theory, where the reduced Planck mass $M_{p}=2.4\times 10^{18}~{\rm GeV}$. 

We are assuming that in 4 dimensions $M_{p}$ puts
a natural cut-off here for any physics beyond the Standard Model.
The another assumption we have made here is that the range of flatness of the potential of inflation
$|\Delta\phi|=|\phi_{cmb}-\phi_{e}|{\sim O}(10^{-1}M_{p})<M_{p}$, for which the model is fully embedded within a particle
theory such as that of gauge invariant flat directions of minimal supersymmetric Standard Model (MSSM), or MSSM$\times U(1)_{B-L}$~\cite{Enqvist:2010vd}.
Here $\phi_{cmb}$ and $\phi_{e}$ are the inflaton field at the Hubble crossing and the end of inflation respectively.

Both these sub-Planckian constraints leads to the observed tension of the low power of the
CMB at low-$l$ with the high-$l$, which leads to a negative curvature of the
power spectrum at small $k$ and a positive curvature of the power
spectrum at large $k$, see~\cite{Planck-1}. For an inflection-point model of inflation this
will provide an improved constraint on PBH formation and GW waves via running in the power spectrum (see fig~(\ref{fig1}) for the details.).

The fractional density of PBH formation can be calculated as:
\be\begin{array}{lll}\label{gam1} \displaystyle f( \geq{M}) =  \gamma~\text{erf}\left[\frac{\varTheta_{\rm th}}
{\sqrt{2} \varSigma_{\varTheta}(k_{PBH})}\right].\end{array}\ee

In general the mass of PBHs is expected to depend
on the amplitude and the shape of the primordial perturbations.
The relation between the PBH formation scale ($k_{PBH}=1/R$) and the PBH mass can be expressed as:
\be\begin{array}{llll} \label{Ras1}
\displaystyle  
k_{PBH}= \frac{\sqrt{\gamma}}{5.54 \times 10^{-24}}\left(
\frac{{M}_{PBH} }{1\ \text{g}} \right)^{-\frac{1}{2}}
\left(\frac{g_{*}}{3.36}\right)^{-\frac{1}{6}}~ Mpc^{-1}\,.\end{array}
\ee
Moreover, we can express the fractional density of 
PBH formation in terms of the PBH abundance at the present epoch, $\eta_{0}$, as \cite{Saito:2009}:
\be\begin{array}{llll}\label{oPBH}
\displaystyle \Omega_{PBH}h^{2}(\eta_0)=10^{14}\times f\left(
\frac {M_{\text{PBH}}} {10^{20}\ \text{g}} \right)^{-\frac{1}{2}}\left(\frac{g_{*}}{3.36}\right)^{-\frac{1}{3}}.
\end{array}\ee
The recent observations from {\it Planck} puts an upper bound on  the amplitude of {\it primordial gravitational waves}
via tensor-to-scalar ratio, $r(k_*)=P_T/P_S$. This bounds the potential energy stored in the inflationary potential, i.e. 
$V_{*}\leq (1.96\times 10^{16}{\rm GeV})^{4}(r(k_{*})/0.12)$~\cite{Ade:2013uln}. 

With the help of Eqs.~(\ref{cfr1},~\ref{dx2},~\ref{sadtc1},~\ref{Ras1},~\ref{oPBH}),
we can link the GW abundance at the present  time:
\be\begin{array}{lll}\label{xca1}
\displaystyle \Omega_{GW}h^{2}(\eta_0)=\left(\frac{\varSigma_{\varTheta}(k_{PBH})}{10^{-2}}\right)^{2}
\left(\frac {M_{\text{PBH}}} {10^{20}\ \text{g}} \right)^{1/2}
\frac{ \Omega_{PBH}h^{2}f^{-1}}{1.7\times 10^{21}}.
\end{array}\ee
In order to realize inflation below the Planck scale, i.e. $M_{p}$, we need to observe the constraint on 
the flatness of the potential, i.e. $\Delta \phi$, via the tensor-to-scalar ratio as recently
 obtained in Refs.~\cite{Choudhury:2014wsa}~\footnote{There was a very minor numerical error in earlier computation in Ref.~\cite{Choudhury:2013iaa,Choudhury:2014kma}, 
which we have now corrected it in Ref.~\cite{Choudhury:2014wsa}. But the final conclusion remains unchanged due to such correction.} :

\be\begin{array}{llll}
\displaystyle\frac{\left |\Delta\phi\right|}{M_p} \approx\frac{6}{25}\sqrt{\frac{r(k_{*})}{0.12}}\left|\frac{r(k_{*})}{16}-\frac{\eta_{V}(k_{*})}{2}-
1+\cdots \right|\,.\label{con10}
   \end{array}\ee
Note that it is possible to saturate the Planck upper bound on tensor-
to-scalar ratio, i.e. $r(k_{*})\leq 0.12$ for small field excursion characterized by, $\phi_0<M_{p}$
and $|\Delta \phi|<M_{p}$. Additionally,  $\cdots$ contain the higher order terms in the slow roll parameters which 
only dominate in the next to leading order.


\begin{figure}[t]
\centering
\includegraphics[width=7.3cm,height=6.5cm]{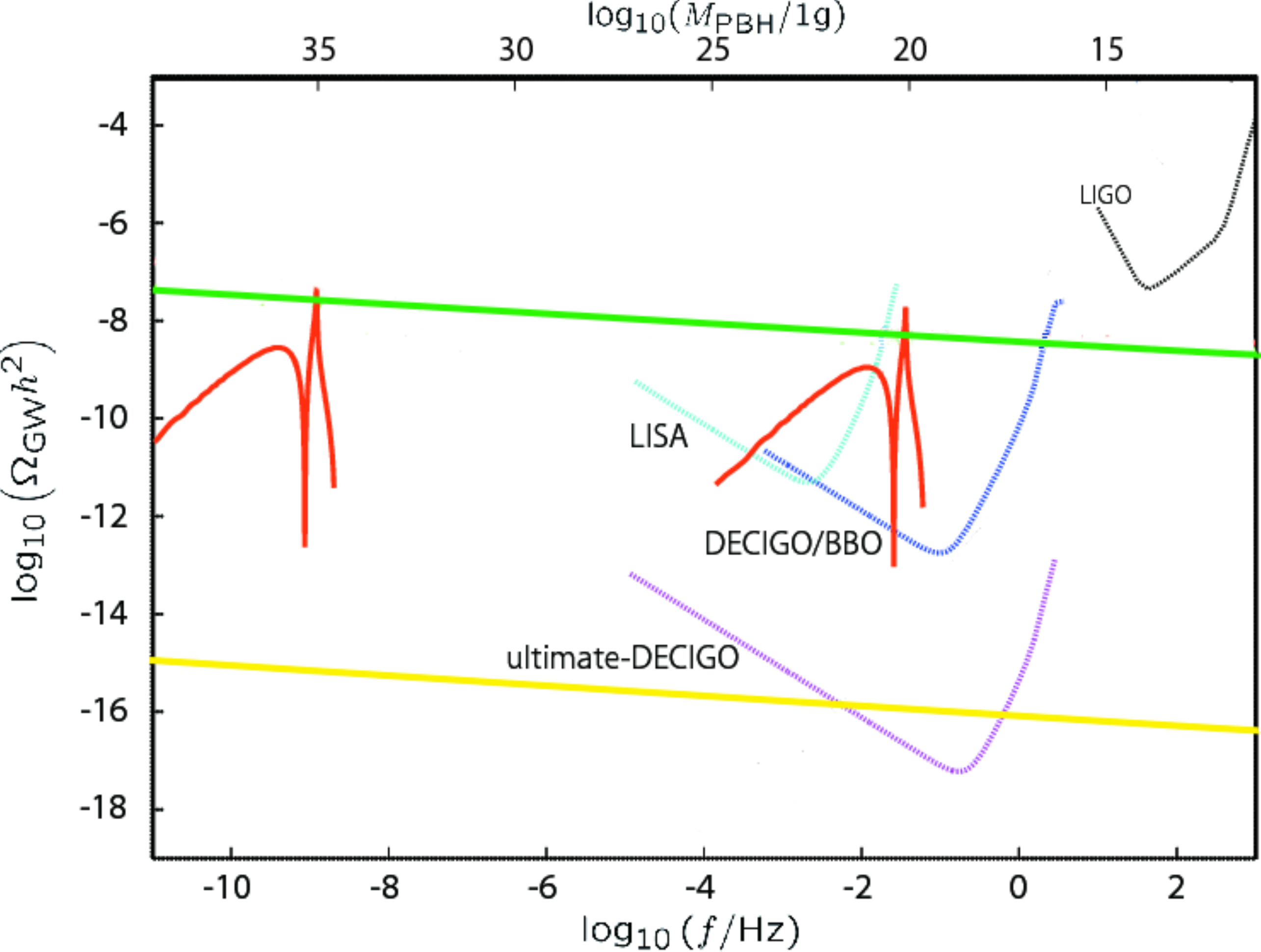}
\caption{\footnotesize $\Omega_{GW}h^{2}$ and $M_{PBH}$ have been plotted in a Logarithmic scale where 
left and right wedge-shaped {\it red} colored curves indicate power spectra of GWs from two different peaked
fluctuations corresponding to the set of values for $(\Omega_{PBH}h^2,~M_{PBH})$=
($10^{-5}$, 30$M_{\odot}$) (left) and ($10^{-1}$, $10^{22}$ g) (right) for relativistic degrees of freedom $g_{*}=228.75$,
 from Eq.~(\ref{oPBH}) and Eq.~(\ref{xca1}). The green and yellow solid line shows an
envelope curve, corresponding to high $V^{1/4}_*=1.96\times 10^{16}$GeV, and low scale $V^{1/4}_*=6.48\times 10^{8}$GeV, sub-Planckian inflationary models
obtained by varying the scale of PBH formation $k_{PBH}$ and the tensor-to-scalar ratio $r$, which depend on the 
frequency of GW in a logarithmic scale. 
}
\label{fig1}
\end{figure}


Collecting the real root of the tensor-to-scalar ratio, $r$, in terms of the field displacement $|\Delta\phi|$ from Eq.~(\ref{con10}),
at the leading order, we can derive a closed constraint relationship between $\Omega_{GW}$ and $\Omega_{PBH}$ at the present epoch,
for {\it inflection point inflationary} models:
\be\begin{array}{llll}\label{xca1c}
    \displaystyle
     \Omega_{GW}h^{2}\leq \frac{6\times 10^{-18}}{\gamma}
\left(\frac {M_{\text{PBH}}} {10^{20}\ \text{g}} \right)^{\frac{1}{2}}\frac{ {\cal O}^{2}_{PBH}\Omega_{PBH}h^{2}}{erf
\left(\frac{{\varTheta}_{th}}{\sqrt{2}{\cal O}_{PBH}}\right)}\,.
   \end{array}\ee
%


\begin{figure}[t]
\centering
\includegraphics[width=7.3cm,height=6.5cm]{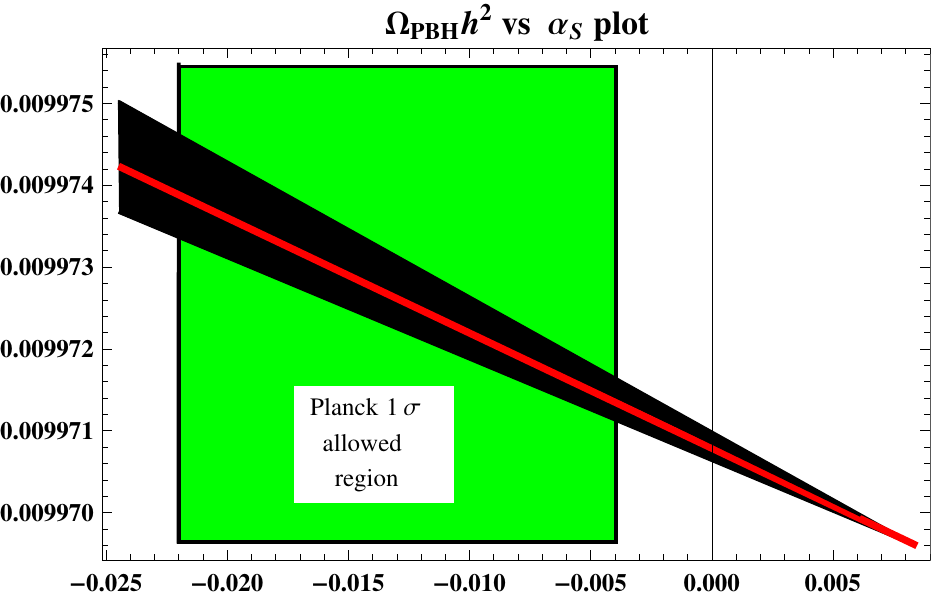}
\caption{\footnotesize We have plotted $\Omega_{PBH}h^{2}$ with respect to the running of the tilt, $\alpha_{s}$. The black colored band shows
Planck allowed region for $n_{s}$ within the range $0.955<n_{s}<0.970$ at $2\sigma$ C.L. within 
the threshold interval $0.2<\varTheta_{\rm th}<0.8$ of {\it Press--Schechter theory}.
The green band for the $1\sigma$ allowed region by Planck from the constraints on the running of the spectral tilt ($\alpha_{s}$) 
within the range $-0.022<\alpha_{s}<-4\times 10^{-3}$. This additionally puts a stringent constraint on the PBH abundance within a region
 $9.99712\times 10^{-3}<\Omega_{PBH}h^{2}<9.99736\times 10^{-3}$. Here the red straight line is drawn for the central value of the the spectral-tilt $n_{s}=0.962$ 
and $\varTheta_{\rm th}=0.5$.
}
\label{fig2}
\end{figure}

where we introduce a new parameter ${\cal O}_{PBH}$, 
 which can be expressed in terms of the inflationary observables as mentioned in the appendix, see Eq.~(\ref{coeff}).
For definiteness,  we also identify the PBH mass with the horizon mass when the peak scale is within the sub-hubble region.
An interesting observation can be made here; typically the PBH mass can also be related to the peak frequency of the GWs 
produced from the collapse of an over-densed region of the space-time to form PBHs, see \cite{Saito:2009}, by following the 
linearized gravitational wave fluctuations up to the second order~\cite{Ananda:2006af}. The peak frequency is given by:
$f^{peak}_{GW}=k_{p}/\sqrt{3}\pi a_{0}$, where $k_{p}$ is the peak value of the momentum scale, and $a_{0}$ is the 
scale factor at the present epoch. Therefore, the final expression yields for frequency and the amplitude~\cite{Saito:2009}:
\be\begin{array}{llll}\label{sq12}
\displaystyle f^{peak}_{GW}=0.03~Hz~\left(\frac{M_{\text{PBH}}}{10^{20}\ \text{g}} \right)^{-{1}/{2}}\left(\frac{g_{*}}{3.36}\right)^{-{1}/{12}}, \\
\displaystyle A^{peak}_{GW}=6\times 10^{-8}\left(\frac{g_{*}}{3.36}\right)^{-{1}/{3}}\left(\frac{\Sigma_{\Theta}(k_{PBH}\approx k_{p})}{10^{-2}}\right)^{2}\,.
\end{array}\ee
where $\Sigma_{\Theta}(k_{PBH}\approx k_{p})$ is explicitly defined in Eq~(\ref{sadtc1}), see Ref.~\cite{Saito:2009}.

It is important to note that the space-based laser interferometers are sensitive to 
GWs with frequency range $10^{-5}{Hz} \lesssim f \lesssim 10 {Hz}$, which 
covers the entire mass range of the PBHs, $10^{20}{\rm g}<M_{{PBH}}<10^{26}{\rm g}$,
which comes from the stringent Dark Matter constraint, $\Omega_{DM}h^{2}=0.1196\pm 0.0031$ from $Planck+WMAP9$ combined dataset
within $1.5\sigma$ C.L \cite{Planck-1}.

LISA \cite{LISA} can probe up to its best sensitivity
 $\Omega_{{GW}}h^2 \sim 10^{-11}$ at GW frequency
$f \sim 10^{-2} {Hz}$ corresponding to the PBH mass $M_{{PBH}} \sim 10^{21}{\rm g}$, 
DECIGO/BBO \cite{BBO} and the ultimate-DECIGO \cite{DECIGO} are designed to probe up t
 $\Omega_{{GW}}h^2 \sim 10^{-13}$ and 
$\Omega_{{GW}}h^2 \sim 10^{-17} $, respectively at the peak frequency 
$f \sim 10^{-1} {Hz}$ with PBH mass $M_{{PBH}}\sim10^{19}{\rm g}$ in its near future run \cite{caltech},~\cite{kudoh:2006}. 
On the other hand the sensitivity of LIGO \cite{LIGO} is too low at present and
in the near future to detect the primordial GWs. 
This implies that for LIGO the abundance of the PBHs 
are constrained within the PBH mass $M_{{PBH}}<7 \times 10^{14}{\rm g}$ with effective GW frequency $f_{{GW}}>10{Hz}$ 
which cannot be observed at the present epoch.

Constraints from all of these GW detectors represented by convex lines with different color codes in Logarithmic scale in Fig.~(\ref{fig1}).
We have also shown the variation of GW abundance for low (green) and high (yellow) scale sub-Planckian models by varying 
PBH mass ($M_{PBH}$) and tensor-to-scalar ratio ($r$) using Eq.~(\ref{xca1c}) and Eq.~(\ref{xca1}) in Fig.~(\ref{fig1}).  
Additionally, we have shown the two wedge-shaped
curves shown in red represented by $(\Omega_{PBH}h^2,M_{PBH})$=
($10^{-5}$, 30$M_{\odot}$) (left) and ($10^{-1}$, $10^{22}$ g) (right) for relativistic degrees of freedom $g_{*}=228.75$.
The appearance of the sharp peaks in the left and right wedge shaped red curves signify the presence of maximum
 value of the GW abundances at the present epoch corresponding to the peak frequency given by Eq.~(\ref{sq12}).
Each wedge shaped curves accompany smooth peaks, this is
 due to the resonant amplification procedure when the peak width for fluctuation, $\Delta_{p}<<k_{p}/2$.
If the peak width exceeds such a limit then the frequency of the fluctuations will increase and 
we get back the peak for sharp fluctuation in the 
right side for each of the wedge shaped curve.

In Fig.~(\ref{fig2}), we have shown the behaviour of the PBH abundance with running of the spectral tilt 
within the Planck $2\sigma$ C.L.(black region) of spectral-tilt~\cite{Ade:2013uln}. We have explicitly 
shown the $1\sigma$ allowed constraint on the running of the spectral tilt by 
the green shaded region which additionally puts a stringent constraint on the PBH abundance 
within a tiny region $9.99712\times 10^{-3}<\Omega_{PBH}h^{2}<9.99736\times 10^{-3}$. 
Note that if we incorporate ${\rm Planck+WMAP9+high~L+BICEP2}$ data \cite{Ade:2014xna}, 
our results would modify, although the physics behind the mechanism 
would remain be the same. We would like to revisit the problem in future with a more detailed study.

\section{Conclusion} 
To summarize, we have shown that it is possible to establish a generic relationship
 between PBH and GW abundance for a sub-Planckian model of inflation with a flat potential, 
 where inflation is driven near an inflection-point.  For such a class of model it is possible to predict 
 $\Omega_{GW}h^2$ and $\Omega_{PBH}h^2$ with the help of this new expression given by Eq.~(\ref{xca1c}).
We have used important constraints arising from various GW detectors, which we have shown in Fig.~(\ref{fig1}),
and the PBH abundance with running of the spectral tilt in Fig.~(\ref{fig2}).

\section{Acknowledgments:} AM would like to thank Andrew Liddle for helpful discussions.
SC thanks Council of Scientific and
Industrial Research, India for financial support through Senior
Research Fellowship (Grant No. 09/093(0132)/2010). AM is supported 
by the Lancaster-Manchester-Sheffield Consortium for Fundamental Physics under STFC grant ST/J000418/1.



\section{Appendix} 

The momentum dependent co-efficients appearing in Eq~(\ref{sadt}) and ${\cal O}_{PBH}$  appearing in Eq~(\ref{xca1c}) are given by:
\begin{widetext}
\be\begin{array}{llll}\label{coeff}
    \displaystyle 
    A=\left[\frac{\sqrt{\pi}}{2}\frac{k_{PBH}}{k_{*}}erf\left[\frac{k}{k_{PBH}}\right]\left(1+\ln\left(\frac{k}{k_{*}}\right)\right)\right]^{k_{\Lambda}}_{k_{L}}+B,\\
   \displaystyle  B=\left[\frac{\sqrt{\pi}}{2}\frac{k_{PBH}}{k_{*}}erf\left[\frac{k}{k_{PBH}}\right]\ln\left(\frac{k}{k_{*}}\right)-
2\left(\frac{k}{k_{*}}\right)~_PF_Q\left[\left\{\frac{1}{2},\frac{1}{2}\right\}~;\left\{\frac{3}{2},\frac{3}{2}\right\}~;
-\frac{k^{2}_{\Lambda}}{k^{2}_{PBH}}\right]\right]^{k_{\Lambda}}_{k_{L}},\\
 \displaystyle C=\left[\frac{\sqrt{\pi}}{4}\frac{k_{PBH}}{k_{*}}erf\left[\frac{k}{k_{PBH}}\right]\ln^{2}\left(\frac{k}{k_{*}}\right)-
\left(\frac{k}{k_{*}}\right)\ln\left(\frac{k}{k_{*}}\right)~_PF_Q\left[\left\{\frac{1}{2},\frac{1}{2}\right\}~;\left\{\frac{3}{2},\frac{3}{2}\right\}~;
-\frac{k^{2}_{\Lambda}}{k^{2}_{PBH}}\right]\right.\\ \left.
\displaystyle ~~~~~~~~~~~~~~~~~~~~~~~~~~~~~~~~~~~~~~~~~~~~~~~~~~~~~~~~~~~~~~~~~~-\left(\frac{k}{k_{*}}\right)~_PF_Q\left[\left\{\frac{1}{2},\frac{1}{2},\frac{1}{2}\right\}~;\left\{\frac{3}{2},\frac{3}{2},\frac{3}{2}\right\}~;
-\frac{k^{2}_{\Lambda}}{k^{2}_{PBH}}\right]\right]^{k_{\Lambda}}_{k_{L}},\\
\displaystyle D=\left[\frac{\sqrt{\pi}}{12}\frac{k_{PBH}}{k_{*}}erf\left[\frac{k}{k_{PBH}}\right]\ln^{3}\left(\frac{k}{k_{*}}\right)-
\left(\frac{k}{2k_{*}}\right)\ln^{2}\left(\frac{k}{k_{*}}\right)~_PF_Q\left[\left\{\frac{1}{2},\frac{1}{2}\right\}~;\left\{\frac{3}{2},\frac{3}{2}\right\}~;
-\frac{k^{2}_{\Lambda}}{k^{2}_{PBH}}\right]\right.\\ \left.
\displaystyle ~~~~~~~~~~~~~~~~~~~~~~~~~~~~~~~~~~~~~~~~~~~~~~~~~~~~~~~~~~~~~~~~~~
-\left(\frac{k}{k_{*}}\right)~_PF_Q\left[\left\{\frac{1}{2},\frac{1}{2},\frac{1}{2}\right\}~;\left\{\frac{3}{2},\frac{3}{2},\frac{3}{2}\right\}~;
-\frac{k^{2}_{\Lambda}}{k^{2}_{PBH}}\right]\right.\\ \left.
\displaystyle ~~~~~~~~~~~~~~~~~~~~~~~~~~~~~~~~~~~~~~~~~~~~~~~~~~~~~~~~~~~~~~~~~~
-
\left(\frac{k}{k_{*}}\right)~_PF_Q\left[\left\{\frac{1}{2},\frac{1}{2},\frac{1}{2},\frac{1}{2}\right\}~;\left\{\frac{3}{2},\frac{3}{2},\frac{3}{2},\frac{3}{2}\right\}~;
-\frac{k^{2}_{\Lambda}}{k^{2}_{PBH}}\right]
\right]^{k_{\Lambda}}_{k_{L}},\\
  \displaystyle  {\cal O}_{PBH}=\frac{5\sqrt{A}(1+w)(8.17\times 10^{-3})^{2}}{12\sqrt{2}\pi
\left(1+\frac{3}{5}w\right)}
\left\{1+\frac{2B\eta_{V}(k_{*})}{5A}
+\frac{B}{500A}\left(\frac{r(k_{*})}{0.12}\right)-\frac{3B}{A}\epsilon_{V}(k_{*})
-\frac{C}{5A}
\xi^{2}_{V}(k_{*})
+\frac{2D}{5A}
\sigma_{V}^{3}(k_{*})
+\cdots\right\}\,
   \end{array}\ee
\end{widetext}
 where $_PF_Q$ represents
generalized Hypergeeometric function.


\end{document}